\def\etal{{\rm et al. }}

\def\kms{{\rm km s^{-1}}}
\newcommand\aap{{\em A}\&{\em A}}

\newcommand\aj{{\em AJ}}
\newcommand\apj{{\em ApJ}}

\newcommand\apjs{{\em ApJS}}

\newcommand\mn{{\em MNRAS}}

\newcommand\nature{{\em Nature}}

\documentclass[runningheads]{aa}  
\usepackage{graphicx}
\usepackage{txfonts}
\usepackage[sort]{natbib}
\bibpunct{(}{)}{;}{a}{}{,}
\begin{document}

   \titlerunning{Effect of bars on the galaxy properties}
   \authorrunning{Vera, Alonso \& Coldwell}

   \title{Effect of bars on the galaxy properties}

   \author{Matias Vera\inst{1},
          Sol Alonso\inst{2}
   		  \and
          Georgina Coldwell\inst{2}
          }

   \institute{ICATE, UNSJ-CONICET, CC49, 5400, San Juan, Argentina \\
              \email{solalonsog@gmail.com}
         \and 
             Facultad de Ciencias Exactas, F\'{\i}sicas y Naturales, UNSJ-CONICET, San Juan, Argentina  
             }
             
   \date{Received xxx; accepted xxx}

  \abstract
   {}
   { With the aim of assessing the effects of bars on disc galaxy properties, we present an analysis of different characteristics of spiral galaxies with strong, weak and without bars.}
   { We identified barred galaxies from the Sloan Digital Sky Survey (SDSS). By visual inspection of SDSS images we classified the face-on spiral galaxies brighter than $g < 16.5$ mag into strong-bar, weak-bar and unbarred. 
With the goal of providing an appropiate quantification of the influence of bars on galaxy properties, we also constructed a suitable control sample of unbarred galaxies with similar redshift, magnitude, morphology, bulge sizes, and local density environment distributions to that of barred galaxies. }
   { We found 522 strong-barred and 770 weak-barred galaxies, which represent a bar fraction of 25.82$\%$, with respect to the full sample of spiral galaxies, in good agreement with several previous studies.
We also found that strong-barred galaxies show lower efficient in star formation activity and older stellar populations (as derived with the $D_{n}(4000)$ spectral index), with respect to weak-barred and unbarred spirals from the control sample. 
In addition, there is a significant excess of strong barred galaxies with red colors.
The color-color and color-magnitude diagrams show that unbarred and weak-barred galaxies are more extended towards the blue zone, while strong-barred disc objects are mostly grouped in the red region. 
Strong barred galaxies present an important excess of high metallicity values, compared to unbarred and weak-barred disc objects, which show similar $12+log\left(O/H\right)$ distributions.  
Regarding the mass-metallicity relation, we found that weak-barred and unbarred galaxies are fitted by similar curves, while strong-barred ones show a curve which falls abruptly,
with more significance in the range of low stellar masses ($log(M_{*}/M_{\sun}) < 10.0$).
These results would indicate that prominent bars produced an accelerating effect on the gas processing,  reflected in the significant changes in the physical properties of the host galaxies.
}
   {}

   \keywords{galaxies: properties - galaxies: spiral - galaxies: {structure}    }

   \maketitle
%

\section{Introduction}

Galactic bars are structures observed in a significant fraction of spiral galaxies and are believed to have an important role in the dynamical evolution of their hosts.
Several simulations show that bars can efficiently transport gas from the outer zones to the innermost central regions of the barred galaxies \citep{wei85,deb98,ath03}.
By interaction with the edges of the bars, the gas clouds suffer shocks producing angular momentum losses and allowing a flow of material toward central kiloparcec scale \citep{SBF90}. 
Moreover, some works show that bars can be destroyed by a large central mass concentration (Roberts et al. 1979; Norman et al. 1996; Sellwood \& Moore 1999; Athanassoula et al. 2005). This finding indicates that currently non-barred disc galaxies possibly had a bar in the past (Kormendy \& Kennicutt 2004), and also, that bars may be recurrent in the galaxy life (Bournaud \& Combes 2002; Berentzen et al. 2004; Gadotti \& Souza 2006).
So, in this context bars formed at different times, and with different conditions, might be present in the barred disc galaxies.

Due to the high efficiency of gas inflow, galactic bars can alter several properties of disc galaxies on relatively short timescales.
In this sence, the presence of bars can affect the star formation activity, stellar population, 
colors, modify the galactic structure \citep{atha83,buta96} and change the chemical composition \citep{comb93,martin95}, contributing to the evolution process of the host galaxies 
(Ellison et al. 2011, Zhou et al. 2015).
In addition, the inflow processes have also been  considered an efficient mechanism for 
trigger active galactic nuclei (AGN)  \citep{comb93,cor03,alonso13,alonso14}, 
and to form bulges or pseudo-bulges (e.g., Combes \& Sanders 1981; Kormendy \& Kennicutt 2004; Debattista et al. 2005, 2006; Martinez-Valpuesta et al. 2006; M\'endez-Abreu et al. 2008; Aguerri et al. 2009).

With respect to the relation between bars and host galaxy colors from statistical analysis, different studies show diverse results.
Several observational works found that the bar fraction, $f_{bar}$, is higher in later-type 
spiral galaxies that are bluer and 
less concentrated systems (e.g. Barazza et al. 2008, Aguerri et al. 2009). 
However, other studies displayed an excess of barred galaxies with redder colors from different samples. \cite{master10a} found a high fraction of bars in passive red spiral galaxies for a sample obtained from the Galaxy Zoo catalog \citep{lintott11}. 
In addition, \cite{oh12} showed that a significant number of barred galaxies are redder than their counterparts of unbarred spiral galaxies.
Recently, in our previous works (Alonso et al. 2013, 2014) we found an excess of red colors 
in spiral barred AGN with respect to unbarred active galaxies in a suitable control sample.

The role of the bars on star formation and metallicity have been the subject of several works, 
showing unclear conclusions.
Many studies found that bars enhanced the star formation rates (SFR) in spiral galaxies compared with unbarred ones (e.g. Hawarden et al. 1986; Devereux 1987; Hummel 1990), while other works show that bars do not guarantee increase in star formation activity (Pompea \& Rieke 1990; Martinet \& Friedli 1997; Chapelon et al 1999).
In the similar way, different authors found diverse results in the metalicity studies in barred galaxies with respect to their unbarred counterparts (e.g. Vila$-$Costas \& Edmunds 1992; Oey \& Kennicutt 1993; Martin \& Roy 1994; 
Zaritsky et al 1994; Considere et al. 2000, Ellison et al. 2011).
More recently, by using data from the CALIFA survey, S\'anchez-Bl\'azquez et al (2014) performed a comparative study of the stellar metallicity and age gradients in a sample of 62 spiral galaxies finding no differences with the presence or absence of bars.

The discrepancy in the results of the bar effects on SFR and metallicity may depend on the host galaxy morphology (Huang et al. 1996; Ho et al. 1997; James et al. 2009) and may be also due to the 
length$/$strength of the bar (Elmegreen \& Elmegreen 1985, 1989; Erwin 2004; Menendez-Delmestre et al. 2007).
Similarly, some studies (e.g. Athanassoula 1992; Friedli et al. 1994; Friedli \& Benz 1995) from numerical simulations found such trends, showing that bar strength is related to the efficiency and quantity of gas inflow, and therefore with the star formation activity and metallicity gradients.

Furthermore, different authors have proposed diverse ways to build control samples from unbarred galaxies, used 
to obtain conclusions from comparative studies, and so the discrepancy in the results could be due to a biased selection of these samples.
In this direction, \cite{perez09} found that a control sample for interacting galaxies should be selected matching, at least, redshift, morphology, stellar masses, and local density environment. This is also a suitable criteria for building control samples of barred galaxies (Alonso et al. 2013, 2014).
Motivated by these finds, in this paper we conducted a detailed analysis of the effect of bars on host galaxy properties, with respect to the unbarred ones by studying different characteristics (e.g. color, stellar population, star formation activity, metallicity) with the aim of assessing whether bar structure in discs play a significant 
role in modifying galaxy properties, and how is this effect.

This paper is structured as follows. Section 2 presents the procedure used to construct 
the catalog of barred galaxies from Sloan Digital Sky Survey (SDSS), the classification 
of the bar structures and the control sample selection criteria.
In section 3, we explore different properties of the barred spirals, in comparison with unbarred galaxies obtained from a suitable control sample.
We analize in details the influence of bars on star formation activity, stellar population, color indexes
 and metallicity in host spiral galaxies, with respect to unbarred ones.   
Finally, section 4 summarizes our main results and conclusions. 
The adopted cosmology through this paper is  
$\Omega = 0.3$, $\Omega_{\lambda} = 0.7$, and $H_0 = 100~ \kms \rm Mpc$.


\section{Catalog of barred galaxies} 

The analysis of this paper is based on the Sloan Digital Sky Server Data Release 7 (SDSS-DR7, Abazajian et al. 2009). 
It uses a 2.5m telescope to get photometric and spectroscopic data which cover near one-quarter
of the celestial sphere and collect spectra of more than one million
objects. DR7 includes 11663 square degrees of sky imaged
in five wave-bands (u,g,r,i,z) containing photometric parameter of
357 million objects. 
The main galaxy sample, which contains about
900000 galaxies with measured spectra and photometry, is essentially
a magnitude-limited spectroscopic sample $r_{lim}<17.77$ (Petrosian magnitude),
and most of galaxies span a redshift range $0<z<0.25$ with a mean
readshift of 0.1 (Strauss et al. 2002).
For this work, several physical properties of galaxies have been derived and published
for the SDSS-DR7 galaxies: gas-phase metallicities, stellar masses, current total and 
specific star-formation rates, concentration index, etc. (Brinchmann et al.
2004; Tremonti et al. 2004; Blanton et al. 2005). These data were
obtained from the MPA/JHU%
\footnote{http://www.mpa-garching.mpg.de/SDSS/DR7/ %
} and the NYU%
\footnote{http://sdss.physics.nyu.edu/vagc/ %
} added-values catalogs.

With the aim of obtaining barred galaxies, we first cross-correlated the SDSS galaxies with the spiral objects obtained from the Galaxy Zoo catalog \citep{lintott11}, which comprises a morphological classification of nearly 900000 galaxies drawn from the SDSS.
In order to cover a wide coverage area, this survey is contributed by hundreds of thousands of volunteers, however due to the large number of classifiers
it becomes complex to maintain an unified criteria.
They define different categories (e.i. elliptical, spiral, merger, uncertain, etc) and give the fraction of votes in each category.
In this study, we selected galaxies that were classified as spiral objects by the Galaxy Zoo team with a fraction of votes $>0.6$. 
Taking this into account, a low fraction of galaxies with non spiral morphological types could be included. 
In adittion, we exclude AGN objects (Coldwell et al. 2014), which could affect our interpretation of the results due to contributions from their emission line spectral features.
Furthermore, as bars are objects which lie on the host disc plane (Sellwood \& Wilkinson 1993) and visual inspection become less efficient while inclination increases, we applied another restriction on the ellipticity of the objects, selecting galaxies with axial ratio $b/a>0.4$.
We also restricted the spiral edge-on galaxy sample in redshift ($z<0.06$) and imposed a 
magnitude cut such as the extinction corrected SDSS $g-$mag is brighter than 16.5. 
With these restrictions, our sample comprises 6771 galaxies, and therefore, 
we can make a plausible visual inspection of a good set of objects.

\subsection{Classification}

We proceeded to select barred galaxies by visual inspection. For this task, 
we used the $g$ + $r$ + $i$ combined color images, obtained from online SDSS-DR7 Image Tool%
\footnote{http://skyserver.sdss.org/dr7/en/tools/chart/list.asp %
}. 
Then, by a detailed visual examination we classified the galaxies into four groups based on the
presence of  bars, taking into account their relative light contribution and length with
respect to the structural properties of the host galaxies.
We can summarize the classification as follows:
522 strong-barred (where the size of the bars have, at least, a 30\% of their host galaxy
sizes), 770 weak-barred (the size of the bars is smaller than 30\% the size of their host
galaxies), 688 ambiguous-barred (objects for which it is difficult to decide whether to have a bar or not), and 3711 non-barred galaxies.
We also found some galaxies with elliptical and irregular appearance which were removed from our sample.
The details of the classification are listed in Table 1, and Fig. 1 shows examples of each galaxy type studied in this work.

Therefore, the final catalogue of the barred galaxies has been constructed with  
1292 spiral objects with strong and weak bars (we excluded ambiguous-barred galaxies that did not 
completely match the bar classification).
This represents a fraction of 25.82$\%$ with respect to the sample of 5003 spiral galaxies with clear classification.
In the same direction, several works carried out by means of visual inspection of different galaxy samples 
(e.g. the RC3 and UGC, \citep{deVau91,nilson73,marinova09,alonso13}) 
finding a bar fraction of 25-30$\%$ in agreement with this work.

\begin{table}
\begin{centering}
\begin{tabular}{|c|c|c|}
\hline 
Galaxy type & N$^0$ of objects & Percentages\tabularnewline
\hline 
\hline 
Non-barred galaxies & 3711 & 54.80\%\tabularnewline
\hline 
Strong-barred galaxies & 522 & 7.71\%\tabularnewline
\hline 
Weak-barred galaxies &  770 & 11.37\%\tabularnewline
\hline 
Ambiguous-barred galaxies & 688 & 10.16\%\tabularnewline
\hline 
Elliptical and irregular galaxies & 1080 & 15.96\%\tabularnewline
\hline 
\hline 
Total sample & 6771 & 100.0 \%\tabularnewline
\hline 
\end{tabular}
\par\end{centering}

\caption{Galaxy classification, numbers and percentages
of objects.}
\end{table}

\begin{figure}
\begin{centering}
\includegraphics[width=85mm,height=100mm ]{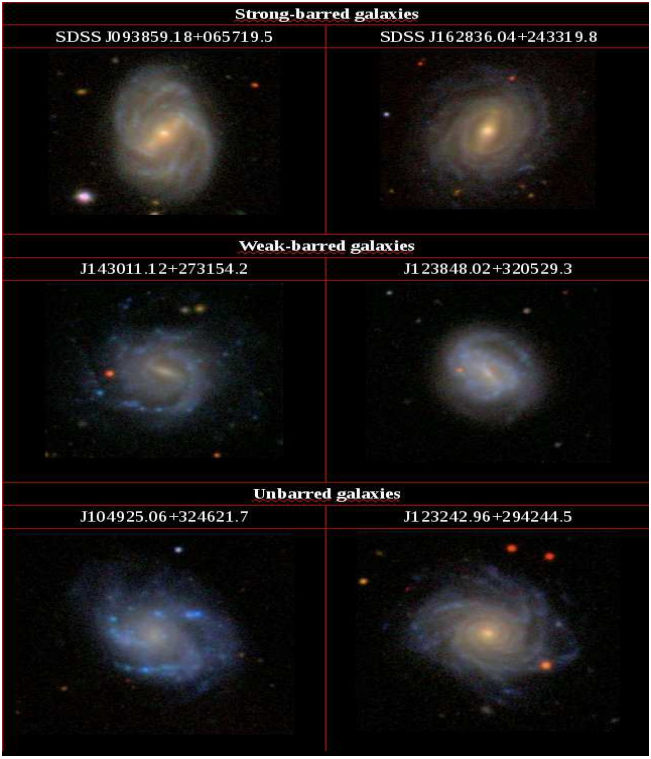}
\par\end{centering}

\caption{Images of typical examples of galaxies classified as strong and weak barred objects (upper and medium panels, respectively). 
Lower panels show examples of spirals without bar.}

\end{figure}

\subsection{Control sample}

To provide a suitable quantification of the impact of bars on the host galaxy properties, we obtained a reliable control sample of unbarred disc objects following Alonso et al. (2013).

We constructed a control sample of galaxies using a Monte Carlo algorithm that selected objects
classified as unbarred galaxies in the previous section with similar redshift and 
extinction and $K-$corrected \citep{blanton03} absolute $r-$band Petrosian magnitude
 distributions of the barred galaxy sample (see panels $a$ and $b$ in Fig. 2).

We have also considered unbarred galaxies in the control sample with similar concentration index ($C$) \footnote{$C=r90/r50$ is the ratio of Petrosian $90\%-50\%$ r-band light radii.} 
distribution to that of the barred catalog to obtain a similar bulge to disk ratio in both samples 
(panel $c$ in Fig.2). 
Furthermore, we restricted the control unbarred spirals to match the $fracdeV$ parameter
defined as the fraction of the light fit by a de Vaucouleurs profile over an exponential profile, where a pure de Vaucouleurs elliptical should have $fracdeV = 1$, and a pure
exponential disc spiral will have $fracdeV = 0$ (Masters et al. 2010a).
Thus this index estimates the bulges surface brigthness distribution, so that it is a good indicator of its size in galaxies with disk morphology (Kuehn 2005, Bernardi et al. 2010, Skibba et al. 2012) 
(see panel $d$ in Fig. 2).
Therefore, the possible differences in the results are associated with the presence of bars and not 
with the difference in the bulge prominences neither with the global galaxy morphology.

In addition, with the aim to obtain galaxies in the same density regions, we also selected objects without bars with similar distribution of the local density environment parameter ($\Sigma_5$) than that of barred galaxies. 
This parameter is calculated through the projected distance $d$ to the 
fifth nearest neighbor galaxy, $\Sigma_5 = 5/(\pi d^2)$, brighter than $M_r = -20.5$ and within a radial velocity difference of less than 1000 km $s^{-1}$ \citep{balo04}.
Fig.2 (panel $e$) shows the distributions of the $log(\Sigma_5)$ for both samples.

With  these restrictions we obtained a control sample of 2205 unbarred spiral objects with similar redshift, brightness, morphology, bulge prominence and local environment to that of barred galaxies. Then, any difference in the galaxy properties is associated only with the presence of the bar, and consequently, comparing the results, we will estimate the real difference between barred and unbarred galaxies, unveiling the effect of this structure on the disc galaxy features.

\begin{figure}
\includegraphics[width=90mm,height=120mm ]{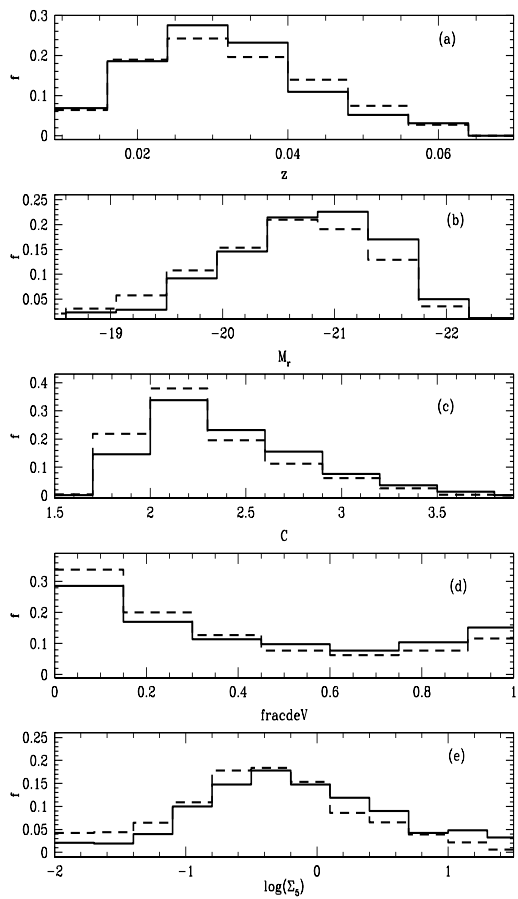}
\caption{Distributions of redshift, luminosities, concentration index, bulge size indicator,  
and local density parameter, 
$z$, $M_r$, $C$, $fracdeV$, and $log(\Sigma_5)$ 
($a$, $b$, $c$, $d$, and $e$ panels), for barred galaxies 
(solid lines) and galaxies without bars in the control sample (dashed lines).}
\label{cont}
\end{figure}


\section{Galaxy properties}

Different studies have shown that bars can induce several processes
that modify many properties of the galaxies (Sellwood \& Wilkinson 1993,
Combes et al. 1993, Zaritzky et al. 1994, Lee et al. 2012, Oh et al. 2012, Alonso et al, 2013, 2014). 
However, there are still many questions about how the properties are modified by the presence of a bar structure in the disc of the spiral galaxies.

In this section we explore the effect of bars, with different structural strength, on the stellar population, star formation activity, colors and metallicity of the host galaxies, in comparison with the unbarred objects in a suitable control sample, obtained from the previous section.
This analysis may help to deepen our understanding of this issue which has been
explored by different authors under diverse approaches.

\subsection{Star formation and stellar population}

With the aim of assesing the effect of bars on the star formation and 
stellar age population, in the following analysis we use the specific star formation rate parameter,
$log\left(SFR/M_{*}\right)$, as a good indicator of the star formation activity in non-AGN galaxies.
It is estimated as a function of the $H\alpha$ line luminosity, and normalized using stellar masses (Brinchmann et al. 2004).  
We also employ the spectral index $D_{n}(4000)$ \citep{kauff03}, that estimates the age of stellar populations. It is calculated from the spectral discontinuity occurring at 4000 $\AA $, produced by an accumulation of a large number of spectral lines in a narrow region of the spectrum, 
specially important in old stars. 
In this analysis, we use the $D_n(4000)$ definition obtained by \cite{balo99}, as the ratio of the average flux densities in the narrow continuum bands (3850-3950 $\r{A}$ and 4000-4100 $\r{A}$). 
The spectroscopic data in SDSS are obtained within the aperture of a spectroscopic fiber (3 arcsec in diameter). 
This corresponds to a typical physical size of the fiber of $\approx$ 2kpc, at the mean redshift of our sample (z $\approx$ 0.03).
Nevertheless, the SDSS spectroscopic parameters (e.g., SFR, metallicity) outside of the fiber are estimated following different methodologies, using the galaxy photometry (see for details, Kauffmann et al. 2003, Tremonti at al. 2004, Brinchmann et al. 2004).

In Fig. 3 (upper panel), we show the star formation activity distributions
for each classified galaxy type. 
It can be clearly appreciated that strong-barred galaxies show a significant excess toward lower $log\left(SFR/M_{*}\right)$ values, with respect to weak-barred and unbarred objects in the control sample. 
In addition, a remarkable bimodality can be observed in galaxies with strong bar prominences. 
This behaviour clearly shows that there is an excess of strong barred galaxies with low star formation activity, indicating that the amount of gas may be not sufficient, after consumed by star formation in the previous process, during prior stages of the galaxy life.
The value located near $log\left(SFR/M_{*}\right)\approx-11.3$ divides both distributions.
Furthermore, weak-barred and unbarred galaxies show similar distribution of specific star 
formation rate. 

Moreover, while several barred galaxies have more concentrated CO in the central region, some early type disks have a lack of CO in this region (e.g. Sheth et al. 2005; Sakamoto et al. 1999).
In this sense, Sheth et al. (2005), using CO observations of the six barred spirals, 
 finding that their sample of barred galaxies have very little molecular gas in the central regions of the galaxies. It could indicate that the gas was consumed in star formation processes at an earlier stage of the galaxy evolution.
More recently, James \& Percival (2016) studied the central regions of four barred galaxies, showing that the star formation activity is observed inhibited within each of these galaxies, suggesting that star formation appears to have been suppressed by the bar.

Lower panel in Fig. 3 shows the distributions of the spectral index $D_{n}(4000)$ for spiral galaxies in the different samples. As we can see, strong-barred galaxies show an important excess toward higher $D_{n}(4000)$ values in comparison with weak-barred and unbarred spiral objects, indicating that strong bars tend to exist in host galaxies with older stellar population. This finding could suggest that: i) strong bars preferentially formed in galaxies with old stellar population or ii) strong bars formed long time ago and thus the stellar population of galaxies became old.
In this direction, Sheth et al. (2008) show that bars were formed first in massive and luminous galaxies, 
and later less massive and bluer systems acquired the majority of their bars. 
Therefore, the strong-bar excess toward higher $D_{n}(4000)$ values could represent bars formed in the first stage, which have grown together with their hosts.
In addition, different authors (e.g., Weinzirl et al. 2009, Laurikainen et al. 2007, 2009) show that, in general, strong bars are more frequently found in early type disk galaxies which are massive than late type objects. 
Similar to that observed in the star formation distributions, strong-barred galaxies show a bimodality in the stellar population around $D_{n}(4000)\approx1.8$. Table 2 quantifies the percentages of galaxies in our different samples with efficient star formation activity and young stellar population. In the similar way, Sanchez-Blazquez et al. (2011) found old stellar population in four barred galaxies. This is in agreement with the findings of James \& Percival (2016) from the spectroscopic analysis, in four different barred disc objects. Besides, old stellar populations have also been found in galaxies with bars by several studies (Perez, Sanchez-Blazquez \& Zurita 2007; Perez, Sanchez-Blazquez \& Zurita 2009; de Lorenzo-Caceres et al. 2012; de Lorenzo-Caceres, Falcon-Barroso, \& Vazdekis 2013), in agreement with our result.

In addition, we divide strong barred galaxies into two subsamples: a group of galaxies that belong to the minor peak of the distributions of both $log\left(SFR/M_{*}\right)$ and $D_{n}(4000)$ (Group 1, G1) and a group of strong barred galaxies belonging to the major peak distributions in Fig. 3 (Group 2, G2). The limits to separate both groups are: $log\left(SFR/M_{*}\right) = -11.3$ and $D_{n}(4000) = 1.8$. 
In this way, we can reveal whether the differences in galaxy properties are mainly driven from having different star formation and stellar population or hosting a bar.

\begin{table}
\begin{centering}
\begin{tabular}{|c|c|c|}
\hline 
Type & $log(SFR/M_{*})>-11.3$ & $D_{n}(4000)<1.8$\tabularnewline
\hline 
\hline 
Strong-barred & 79\% & 77\%\tabularnewline
\hline 
Weak-barred & 97\% & 97\%\tabularnewline
\hline 
Unbarred & 93\% & 91\%\tabularnewline
\hline 
\end{tabular}
\par\end{centering}

\caption{Percentages of spiral galaxies with strong, weak and without bars with $log(SFR/M_{*})>-11.3$ and $D_{n}(4000)<1.8$.}

\end{table}

\begin{figure}
\begin{centering}
\includegraphics[scale=0.38]{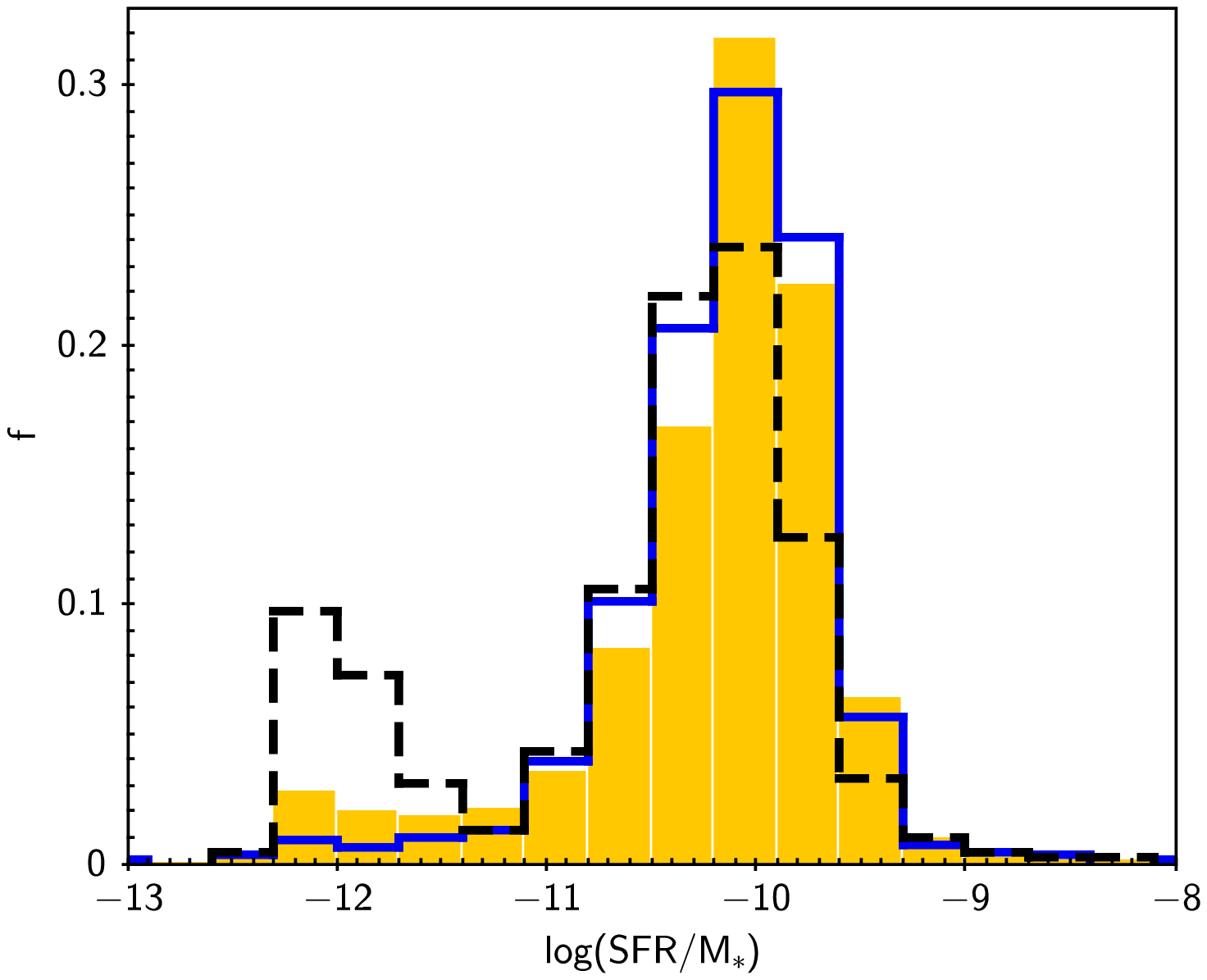}
\par\end{centering}

\begin{centering}
\includegraphics[scale=0.38]{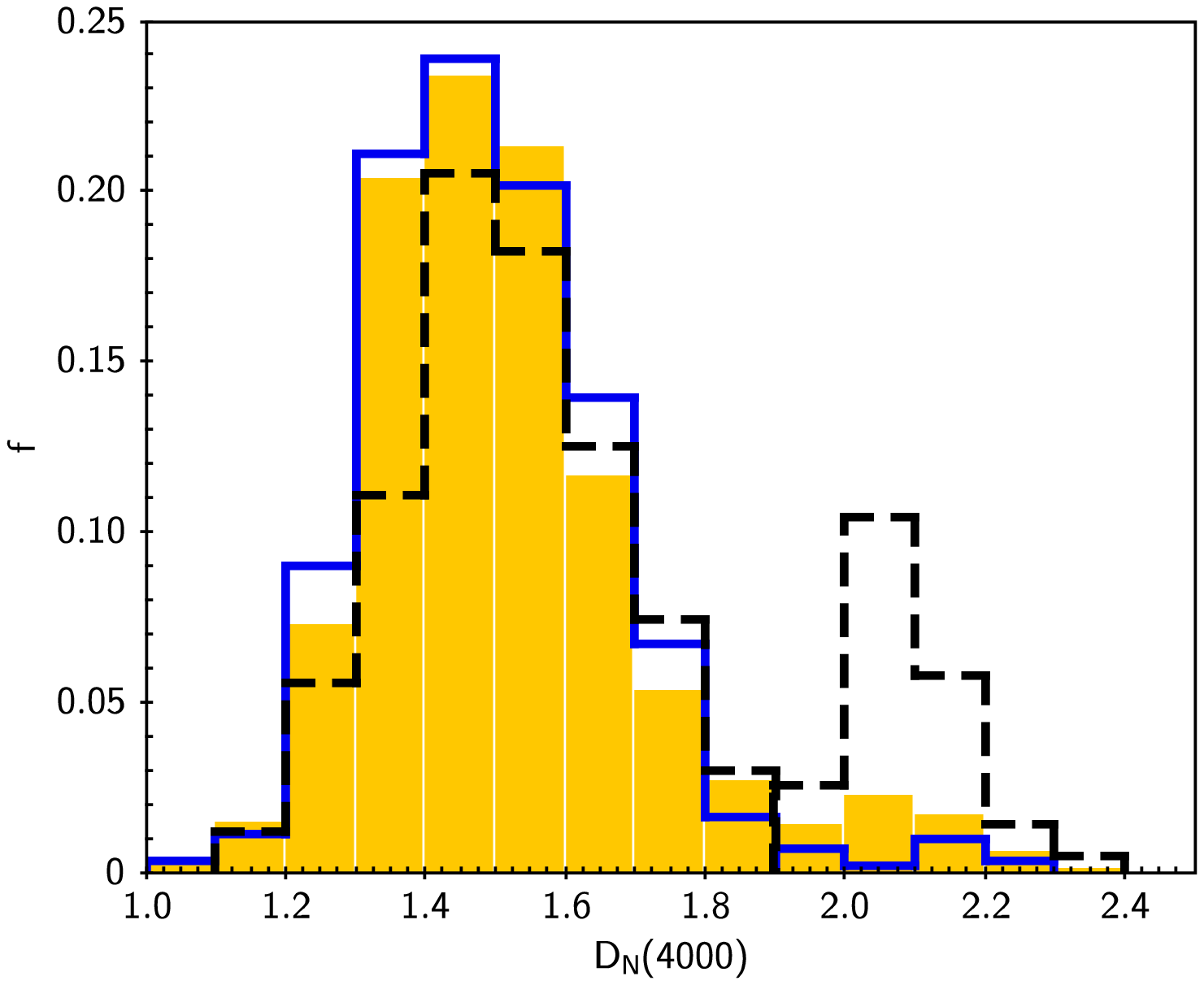}
\par\end{centering}

\caption{Specific star formation rate, $log\left(SFR/M_{*}\right)$ (upper panel) and 
stellar population, $D_{n}(4000)$ (lower panel)
distributions for different galaxy types: unbarred objects (shaded histograms),
weak-barred (solid lines) and strong-barred (dashed lines) galaxies.}
\end{figure}

We checked disc galaxies in the smaller peak in the previous bimodal distributions with older stellar population and low efficient in star formation activity.
We noticed that there are 97 strong-barred lenticular galaxies. In Fig. 4 some examples can be seen. Therefore, an important fraction (about the 20\%) of strong-barred galaxies are SB0 morphological types. 
It could be indicating that usually, when a lenticular galaxy contains a bar, it is an strong structure. These results are consistent with those obtained by Aguerri et al. (2009), who found that bar length (normalised by the galaxy size) in lenticular galaxies tend to be longer than ones in late-types objects. 
Furthermore, our finding agrees with Laurikainen et al. (2009), who found that prominent bars (which measures are calculated using the maximum m$=$2 Fourier density amplitude) are more common in lenticular galaxies than weak bars (considering that lenticular galaxies in their work are sub-divided in S0 and S0/a types, and that a medium amplitude is defined between strong and weak structures).

\begin{figure}
\begin{centering}
\includegraphics[scale=0.17]{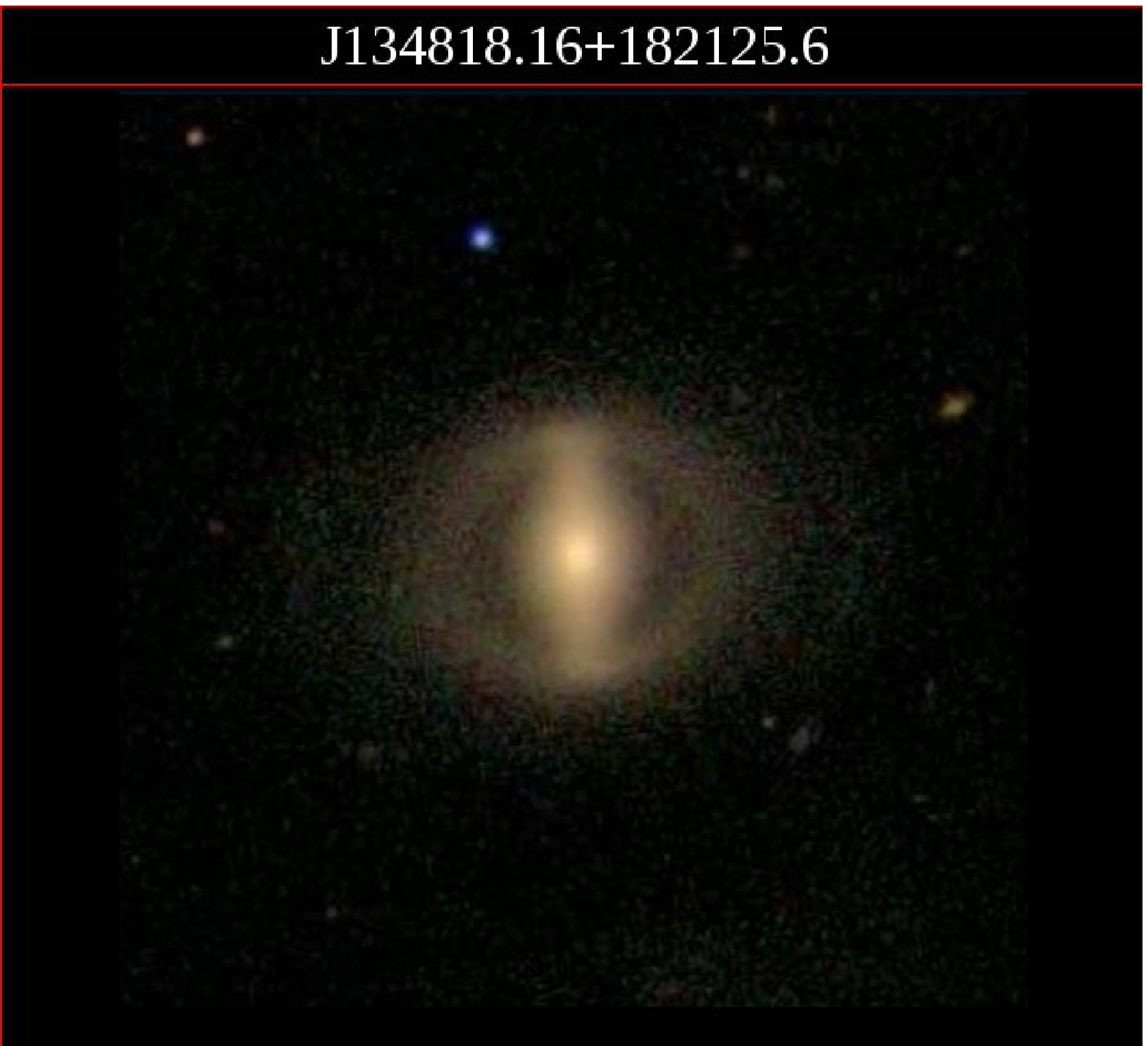} \includegraphics[scale=0.17]{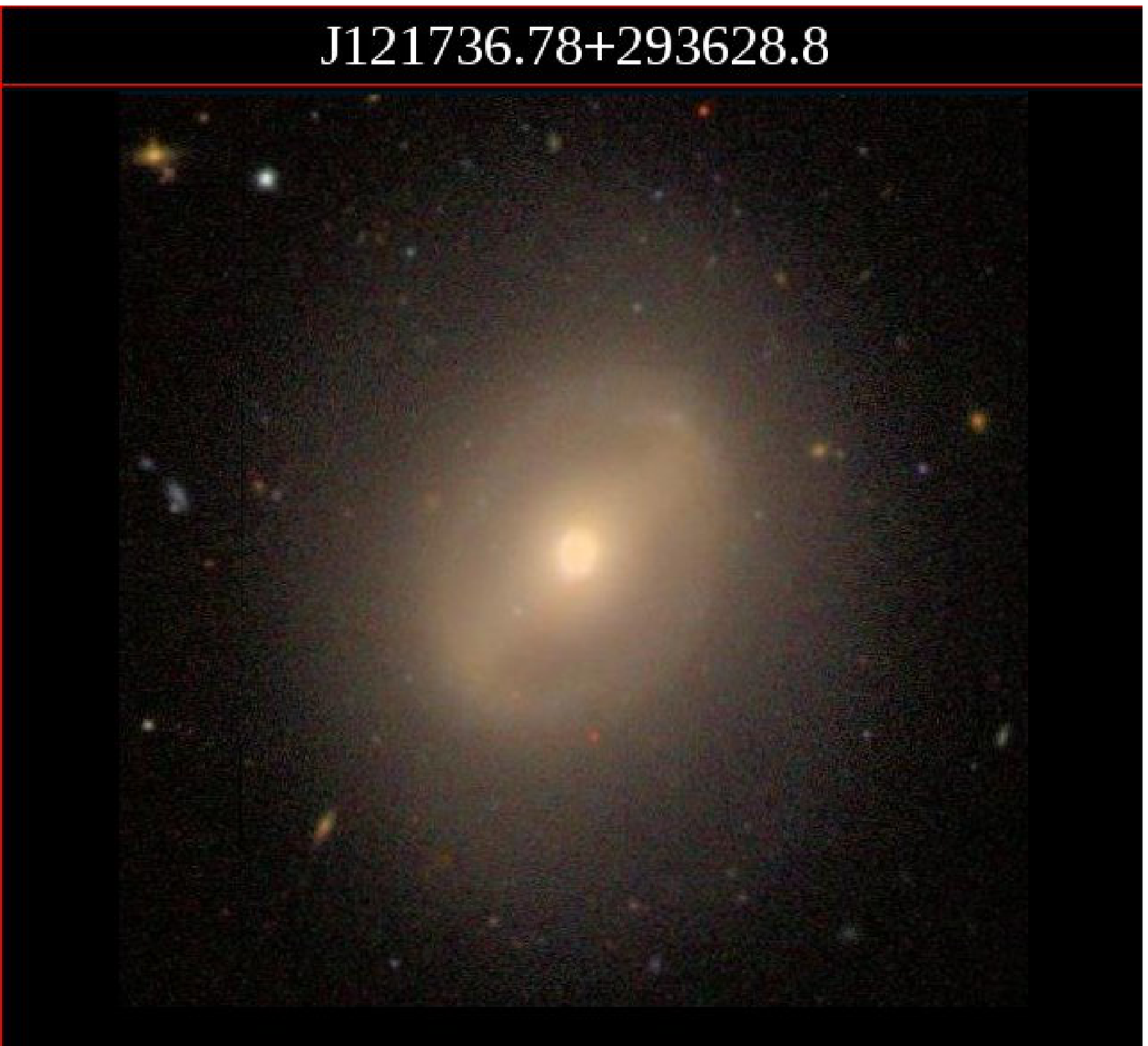} \includegraphics[scale=0.17]{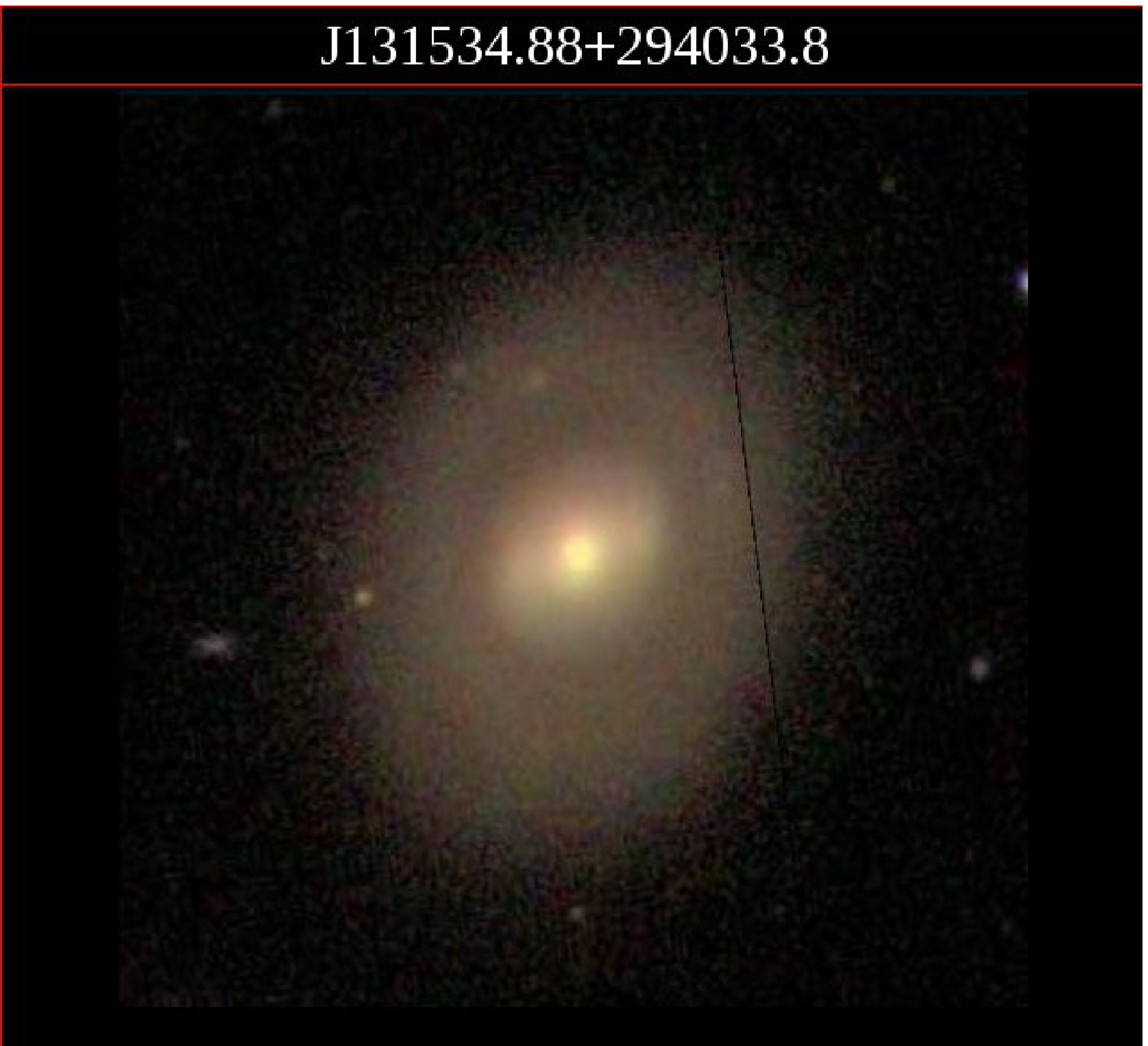}
\par\end{centering}
\caption{Images of three examples of barred lenticular galaxies with 
 $log(SFR/M_{*})<-11.3$ and $D_{n}(4000)>1.8$. 
}
\end{figure}

In order to understand the behaviour of star formation and stellar populations of spirals with strong, weak and without bars, with respect to the stellar masses and the morphology of the host galaxy, we have analysed $log\left(SFR/M_{*}\right)$ and $D_{n}(4000)$, as a function of $log(M_{*})$ and concentrarion index, $C$.
Fig. 5 shows the mean $log\left(SFR/M_{*}\right)$ and $D_{n}(4000)$ as a function of the
stellar mass. Errors were estimated by applying the bootstrap resampling technique in all figures (Barrow et al. 1984).
 For the Fig. 5 and 6 we considered barred galaxies that belong to the major peaks in Fig. 3, with the aim to exclude the fraction of the lenticular galaxies. 
As can be seen, star formation activity decreses towards higher stellar masses 
and, in the same direction, young stellar population diminish with $log (M_{*})$. 
 Clearly host galaxies with strong bars show a systematically lower efficient of star formation activity and older stellar population in all stellar mass bins, with respect to the other samples studied in this paper.
Furthermore, disc objects in the control sample show efficient activity in star formation and younger stellar population. In addition, the trends for weak barred galaxies are observed between strong barred objects and control sample.

\begin{figure}
\begin{centering}
\includegraphics[width=60mm,height=50mm ]{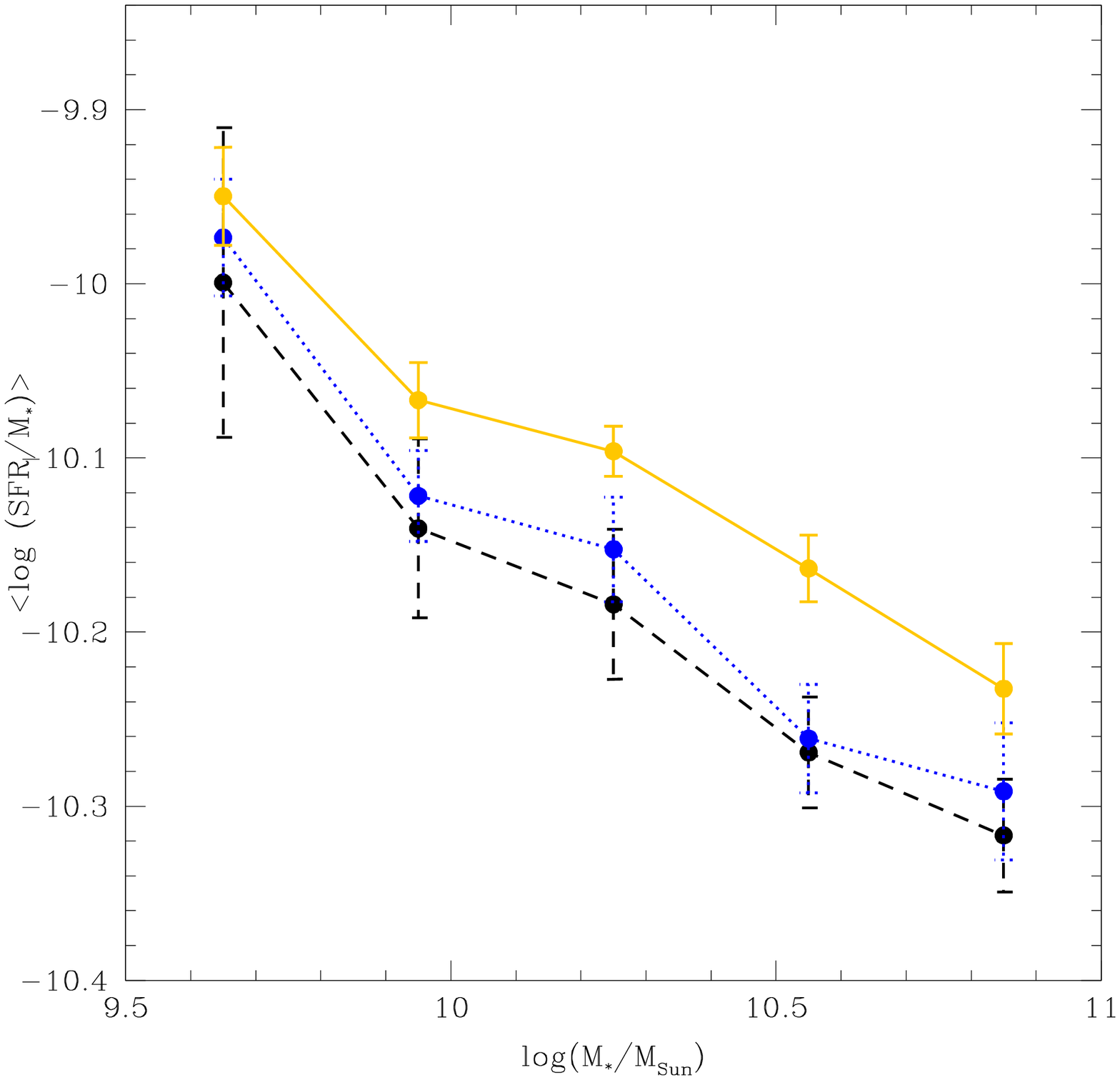}
\par\end{centering}

\begin{centering}
\includegraphics[width=60mm,height=50mm ]{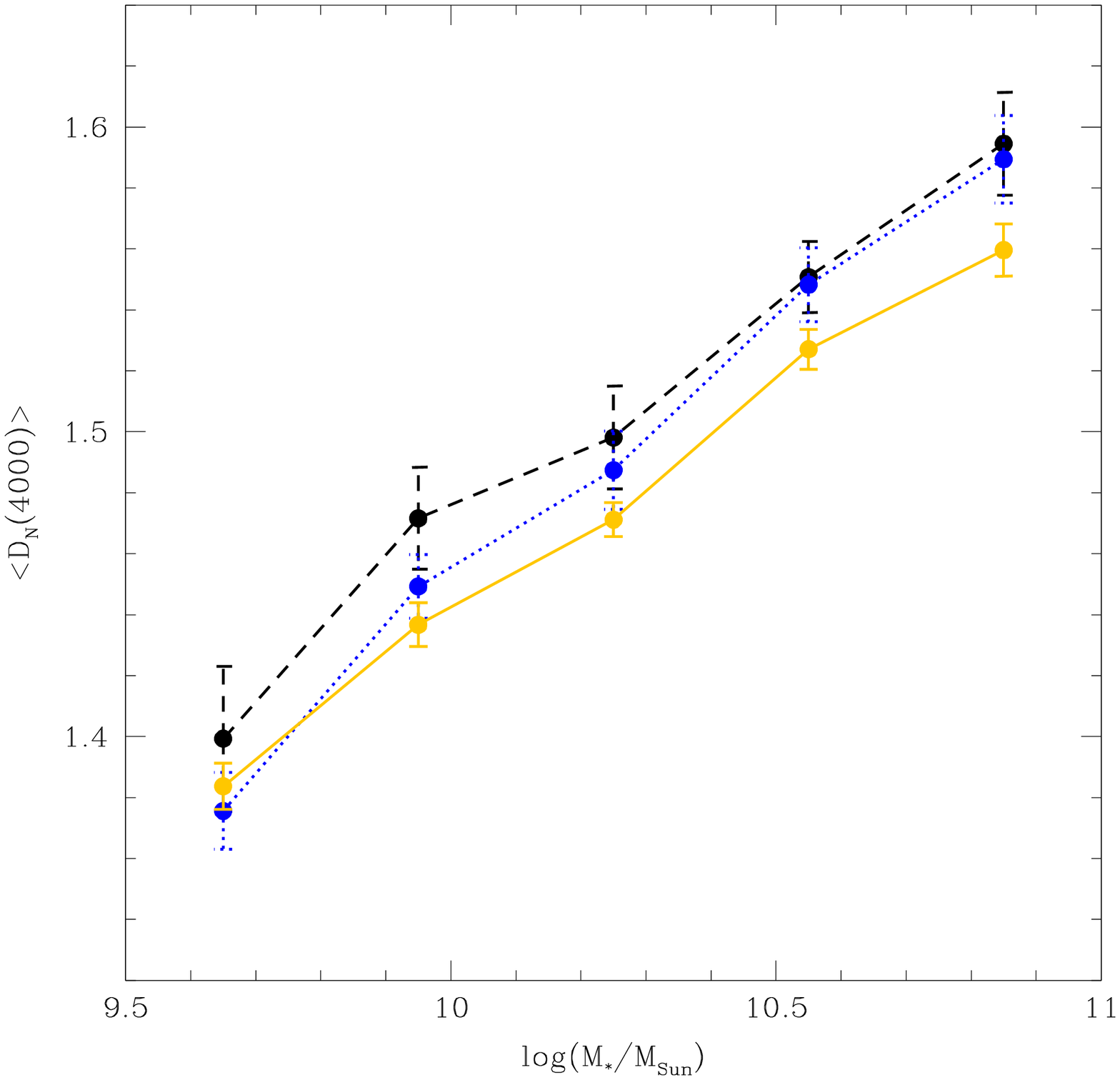}
\par\end{centering}

\caption{Specific star formation rate, $log\left(SFR/M_{*}\right)$, and $D_{n}(4000)$ 
(upper and lower panels) as a function of stellar mass bins
for unbarred, weak and strong barred spiral galaxies that belong to the major peaks of the star formation and stellar population distributions (see Fig. 3) (solid, dotted and dashed lines, respectively).}

\end{figure}

In addition, in Fig. 6 we can see clearly that strong barred galaxies become less efficient star 
formers while increases the $C$ index.  
On the other hand, for unbarred and weak-barred galaxies $log\left(SFR/M_{*}\right)$ remains almost constant for the all $C$ values. 
Moreover, strong-barred objects show older stellar population towards galaxies with earlier morphology.
 Unbarred spirals show younger stellar age population and higher SFR values, for different morphological types, while weak barred galaxies present intermediate tendencies.
This fact could be indicating that bars tend to modify quickly their host galaxy properties (i.e., bars could accelerate the gas processing), when they have became prominent enough.

 In the same direction, Masters et al. (2012) found a significantly lower bar fraction on 
the gas-rich galaxies with respect to gas-poor disc objects. 
The authors suggest that in gas-rich discs the bars funnelling the gas into the central 
region of the galaxy. Then, this material can turned into molecular gas and eventually trigger star formation acticity (e.g. Ho et al. 1997; Sheth et al. 2005; Ellison et al. 2011; Lee et al. 2012). 
So, this process could accelerate the gas consumption, ceasing the formation of the new stars by removing gas from the outer regions of the disk and become red host galaxies.
These mechanisms may indicate different evolutionary stages of the bars in spiral galaxies, which depend on the strength of the bar structure.
In this context, Jogee et al. (2005), based on the properties of circumnuclear gas and star formation, proposed a possible scenario of bar-driven dynamical evolution of the galaxies.
In the first phases, large amounts of gas are transported by the bars towards the galactic central regions, along with an efficient star formation activity.
Then, in the poststarburst phase, the gas has been consumed by circumnuclear starburst, showing low SFR (Sheth et al. 2005). 
 In this context, we argue that during these different stages, the length/strength of the bars and host galaxy properties are modified.

\begin{figure}
\begin{centering}
\includegraphics[width=60mm,height=50mm ]{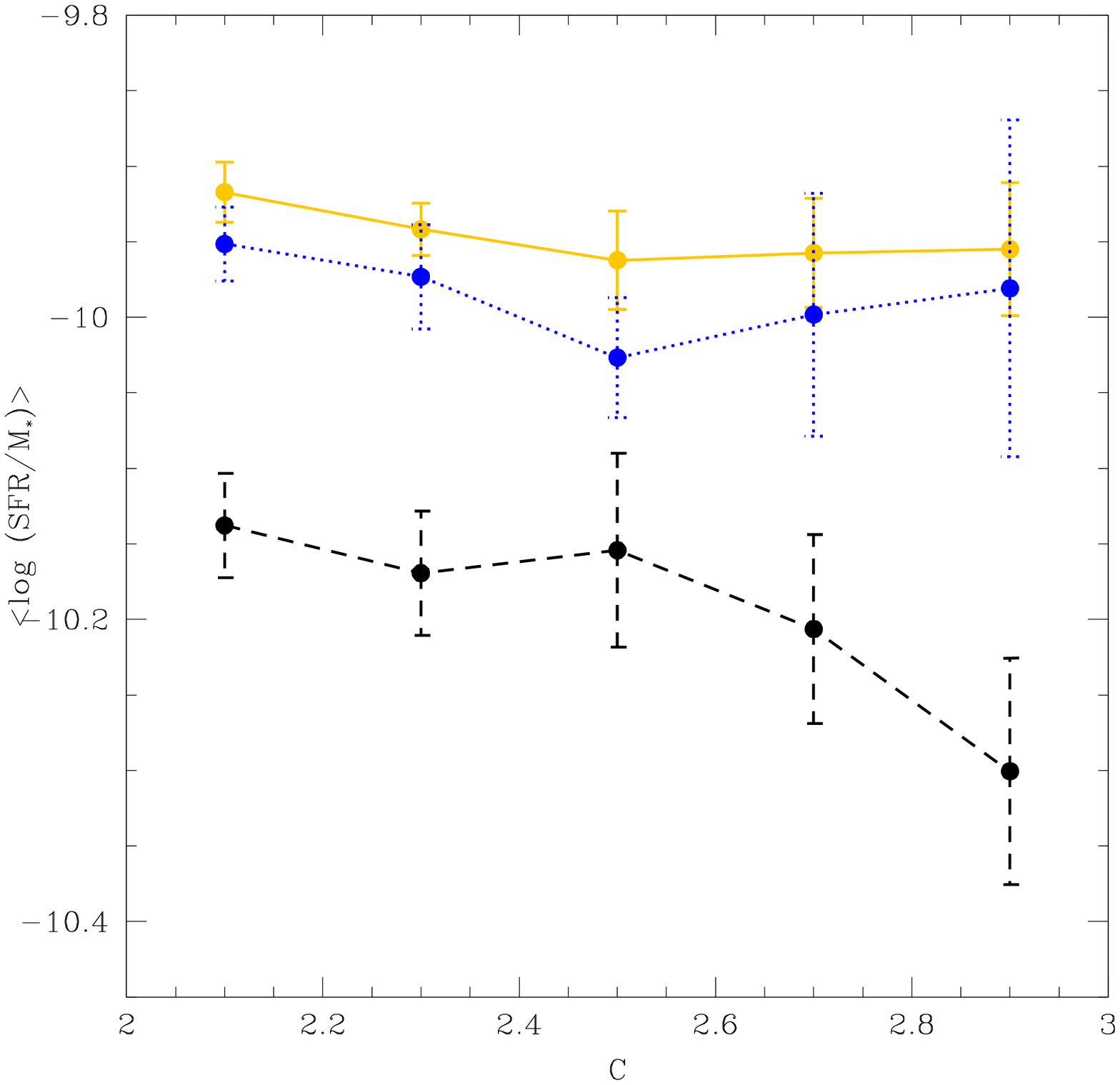}
\par\end{centering}

\begin{centering}
\includegraphics[width=60mm,height=50mm ]{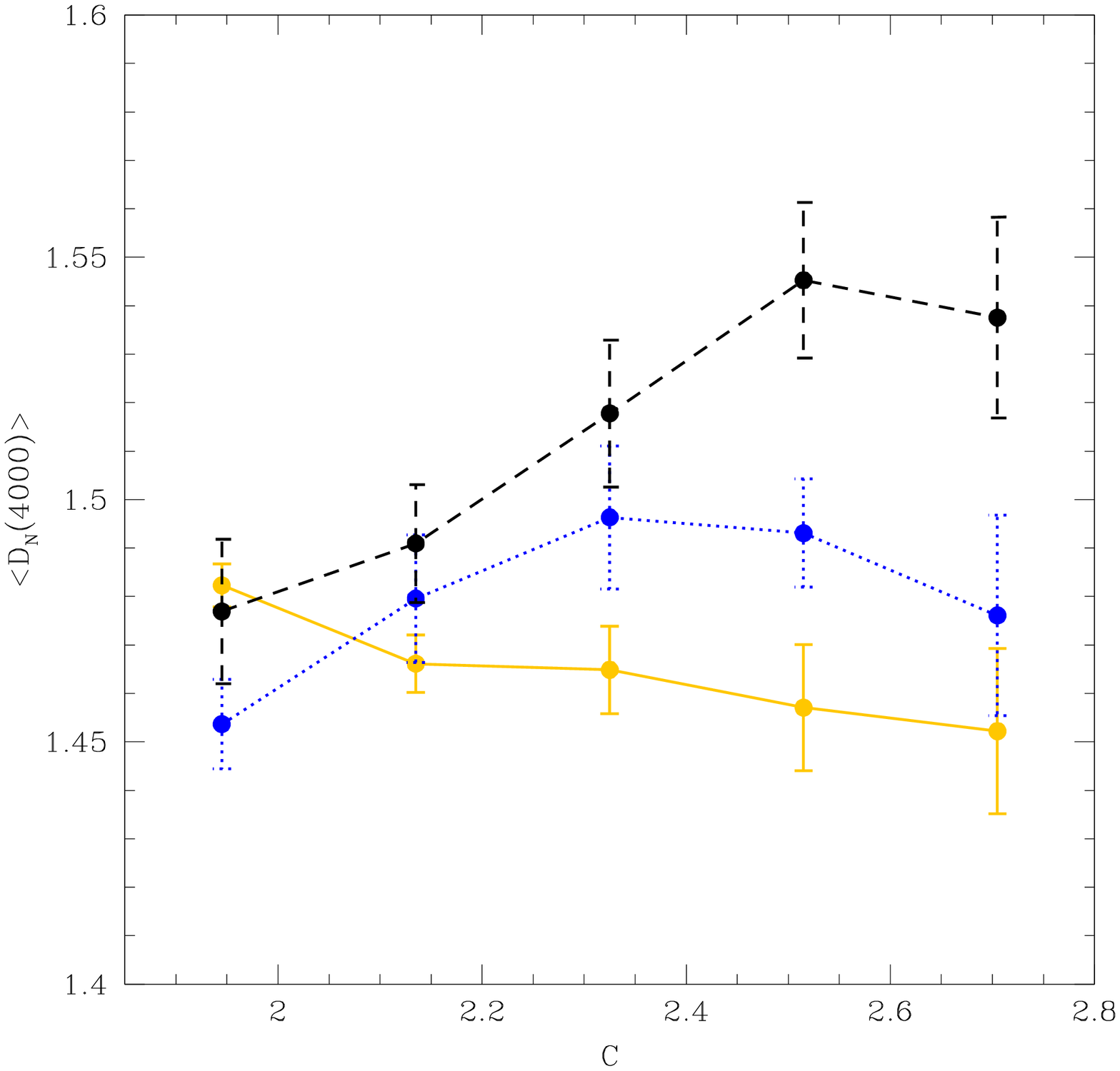}
\par\end{centering}

\caption{Specific star formation rate ($log\left(SFR/M_{*}\right)$, upper panel) and stellar age population ($D_{n}(4000)$, lower panel)
 as a function of $C$ parameter for unbarred, weak-barred
and strong-barred galaxies that belong to the major peaks of the star formation and stellar population distributions (see Fig. 3) (solid, dotted and dashed lines, respectively).}

\end{figure}

\subsection{Colors}

With the aim to explore the colors of the barred galaxies with different structural strength, in Fig. 7 we illustrate the $\left(u-r\right)$ and $\left(g-r\right)$ color distributions for different galaxy types classified previously. 
Strong-barred objects show a clear excess of redder colors, while unbarred and weak-barred galaxies have similar ($u-r$) and ($g-r$) distributions. 
In particular, strong bars in the Group 1 show a significant fraction of host galaxies with extremely 
red colors ($u - r > 2.0$ and $g - r > 0.7$). 
This finding could be reflected low efficiency in star formation activity, old stellar population and earlier morphological galaxy types.
In the same direction, Masters et al. (2011) found that red spiral galaxies have a higher fraction of bars than that in the blue ones, in a sample obtained from Galaxy Zoo catalogue. 
Similarly, Oh et al. (2012) and Alonso et al. (2013, 2014) observed the same behaviour 
for AGN barred hosts. 
Therefore, it seems that bars play an important role in the modification of the host galaxy colors, but only when this structure has become prominent enough.

\begin{figure}
\begin{centering}
\includegraphics[scale=0.38]{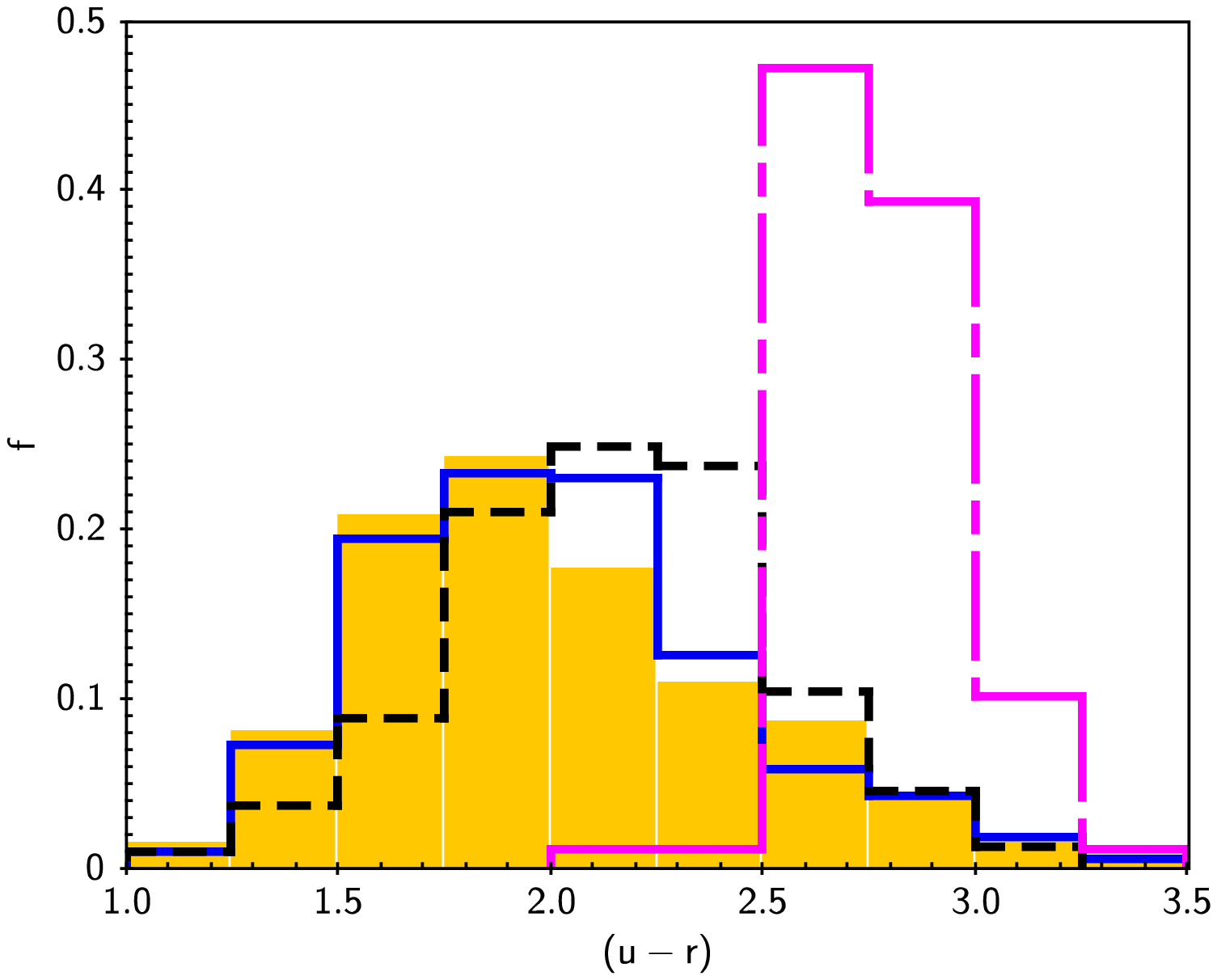}
\par\end{centering}

\centering{}\includegraphics[scale=0.38]{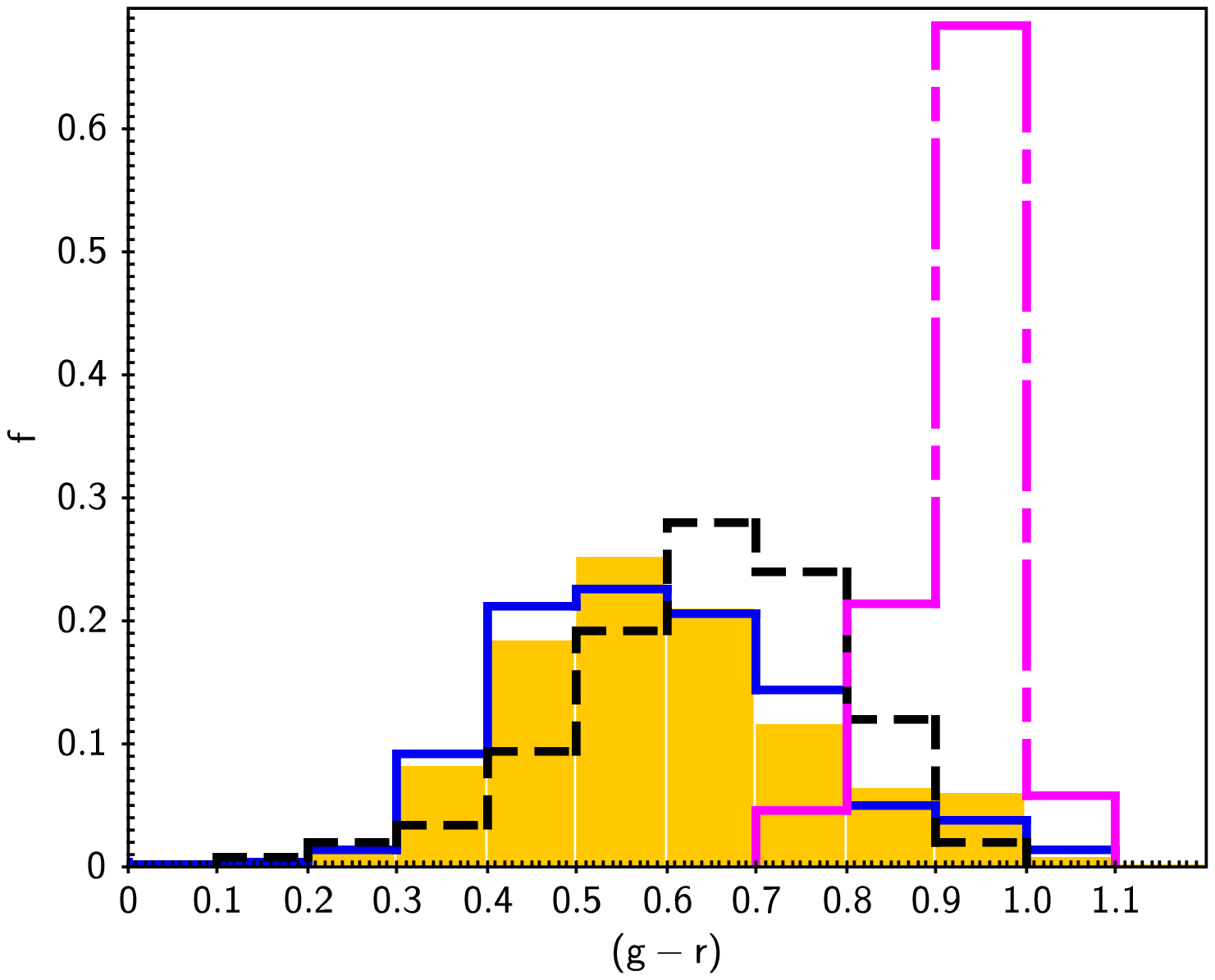}
\caption{Color distributions $\left(u-r\right)$ (upper panel) and $\left(g-r\right)$
(lower panel) for unbarred (shaded histograms), weak-barred (solid lines) and strong-barred  galaxies  in Group 1 (dot-dashed lines) and in Group 2 (dashed lines).}
\end{figure}

In addition, in Fig. 8 we present color-magnitude diagram for the different studied
galaxy types. It can be seen that strong-barred galaxies are principally 
concentrated in the top area (red color region), while unbarred and weak-barred objects
are more uniformly distributed. 
 It is clear that galaxies with strong bars in the Group 1 are located in the redder region of the 
color-magnitude diagram with respect to galaxies from the other samples.
We have also plotted the color fit, developed 
by Masters et al. (2010b), which separate blue and red populations ($(g-r) = 0.67-0.02\left[M_{r}+22\right]$). 
As we can see, galaxies with strong bars are mostly located above the line, while unbarred and 
weak-barred objects lies mostly under the line (blue color region). 
This findings could indicate that, at the same magnitude, strong barred galaxies are usually redder objects with respect to the other samples studied in this work.
Table 3 quantifies the percentage of different galaxy types located in the red color region.

\begin{table}
\begin{centering}
\begin{tabular}{|c|c|}
\hline 
Galaxies & $(g-r)\geq0.67-0.02\,(M_{r}+22)$\tabularnewline
\hline 
\hline 
Unbarred & 34.82\%\tabularnewline
\hline 
Weak-barred & 33.50\%\tabularnewline
\hline 
Strong-barred G1 & 100.00\% \tabularnewline
\hline 
Strong-barred G2 & 53.50\% \tabularnewline
\hline 
\end{tabular}
\par\end{centering}

\caption{Percentages of galaxies located above the line fitted by Masters et
al. (2010b). }

\end{table}

\begin{figure}
\begin{centering}
\includegraphics[scale=0.30]{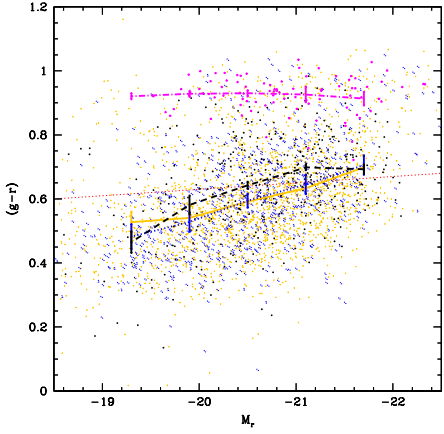}

\caption{Color-magnitude diagram for unbarred (orange crosses), weak-barred (blue open circles) and strong-barred galaxies in Group 1 (magenta stars) and in Group 2 (black fulled circles)
The lines represent the medium values for the different samples: unbarred objects (solid line), weak-barred and strong-barred galaxies (in Group 1 and Group 2) (dotted, dot-dashed and dashed lines, respectively)
Red dashed line plots the fit obtained by Masters et al. (2010b).
}

\par\end{centering}
\end{figure}

Moreover, Fig. 9 illustrates color-color diagram for unbarred, weak- and strong-barred galaxies in both groups. We can note that the three galaxy types lie on a same straight line, although strong-barred ones are mostly grouped in the red region of the diagram
 (mainly those belonging to G1), while the other types show more dispersion 
and are more extended towards the blue region. 
This configuration could be indicating an evolutive relation between the different classified galaxy types. In the first time unbarred disc galaxies could start forming a bar in its central region, from the instability in the disc. This bar would become increasingly prominent while it consumes gas from the disc. 
In the beginning, the bar would not be able to modify significantly the host galaxy characteristics. 
Then, when it reaches a strong prominence, it could start to affect the host galaxy, producing an ascent in the color-color diagram, inducing many important changes in the host galaxy properties.

\begin{figure}
\begin{centering}
\includegraphics[scale=0.30]{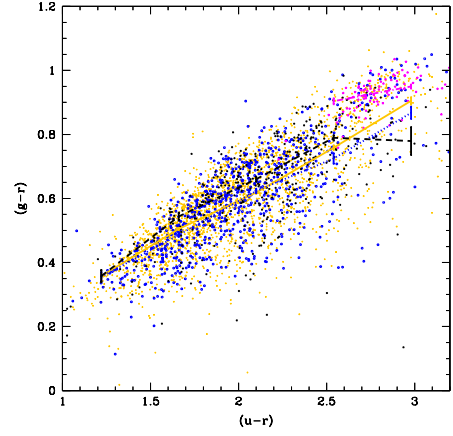}
\caption{Color-color diagram for unbarred disc objects (orange crosses), 
weak-barred (blue open circles) and strong-barred galaxies in Group 1 (magenta stars) and in Group 2 (black fulled circles)
The lines represent the medium values for unbarred objects (solid line), weak-barred and strong-barred galaxies (in Group 1 and Group 2) (dotted, dot-dashed and dashed lines, respectively).
}

\par\end{centering}

\end{figure}

Fig. 10 shows the mean $(g-r)$ and $(u-r)$ colors as a function of the concentration index, 
$C$. It is clear that red objects increase towards the more concentrated galaxies, for the different samples.
This result is consistent with expectations, since galaxies with higher values of the concentration index are related to the bulge type morphology, and lower concentration objects to spiral galaxies. 
 On the other hand, galaxies by evolving passively can become red without increasing concentration parameter, and also some galaxies can increase $C$ without becoming red (e.g. by merging events).
It can be also seen that strong-barred galaxies are redder, for the whole range of the $C$ parameter, compared to their counterparts of weak-barred objects and unbarred hosts in the control sample.
 We notice that this tendency is clearly more significant in galaxies with strong bars that belong to Group 1.
In this context, the results could indicate that galaxies with strong bars could become redder than their counterparts of unbarred and weak-barred disc objects. This fact also supports the idea that intense bars accelerate the gas processing and it is reflected in a reddened population.

\begin{figure}
\begin{centering}
\includegraphics[scale=0.37]{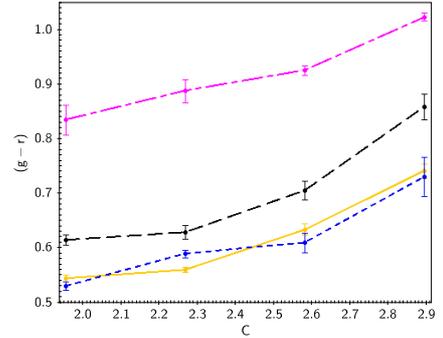}
\par\end{centering}

\begin{centering}
\includegraphics[scale=0.37]{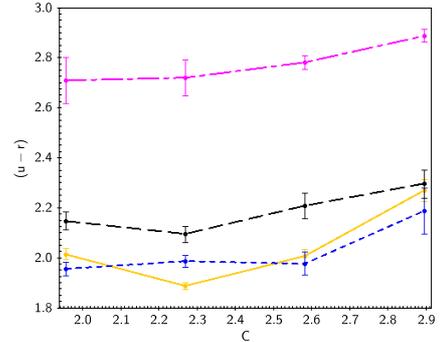}
\par\end{centering}

\caption{Colors, ($g-r$) (upper panel), and ($u-r$) (lower panel), as a function of $C$ parameter, for unbarred, weak-barred and strong-barred (in Group 1 and Group 2) galaxies (solid, dotted, dot-dashed and dashed lines, respectively). }

\end{figure}

\subsection{Metallicity}

The chemical features of the galaxies can store fossil records of their history of formation
since they are the result of diverse physical mechanisms acting at different stages of evolution
(Freeman \& Bland-Hawthorn 2002).
In this sense, metallicity is one of the fundamental physical properties of galaxies, which 
 principally reflects the amount of gas reprocessed by the stars.
In addition, it depends strongly on the evolutive state of a galaxy so it is a good indicator of its age. 
In this analysis, as metallicity parameter, we used $12+log\left(O/H\right)$ which represents
the ratio between oxygen and hydrogen abundances (Tremonti et al. 2004). 
 We found that $\approx$ 80$\%$ of the objects in our samples have $12+log\left(O/H\right)$ measurement, and there is a null fraction of strong barred galaxies in G1 with this parameter.
Therefore, in this section, the sample of the strong-barred galaxies belong to the Group 2.

The influence of the bars in the metallicity can be seen in Fig. 11, in the histograms of the 
$12+log\left(O/H\right)$ for disc galaxies with strong, weak and without bars.  
We also define the low and high metallicity galaxies by selecting two ranges of $12+log\left(O/H\right)$ 
values to have equal number of objects in the control sample. This threshold is 
$12+log\left(O/H\right) = 9.05$. 
From this figure, it can be appreciated that strong-barred galaxies present an important excess 
towards high metallicity values, while unbarred and weak-barred objects show similar distributions.
Table 4 quantifies the excess of disc objects with high metallicity for the different samples. 
This result supports the previous ones, meaning that strong-barred galaxies show low star formation activity, with older/redder stellar populations and higher gas metallicity than weak-barred and 
unbarred spiral galaxies. 
From the chemodynamical simulation studies, Martel et al. (2013) found that the chemical evolution observed within the central region of the disc galaxies depends critically of the prominence of the bar, which evolves strongly with time.

\begin{table}

\begin{centering}
\begin{tabular}{|c|c|}
\hline 
Galaxy Type & $12+log\left(O/H\right)>9.05$\tabularnewline
\hline 
\hline 
Strong-barred & 64.1\%\tabularnewline
\hline 
Weak-barred & 51.4\%\tabularnewline
\hline 
Unbarred & 50.0\%\tabularnewline
\hline 
\end{tabular}\caption{Fraction of galaxies with metallicites higher than 9.05.}

\par\end{centering}

\end{table}

\begin{figure}
\begin{centering}
\includegraphics[scale=0.44]{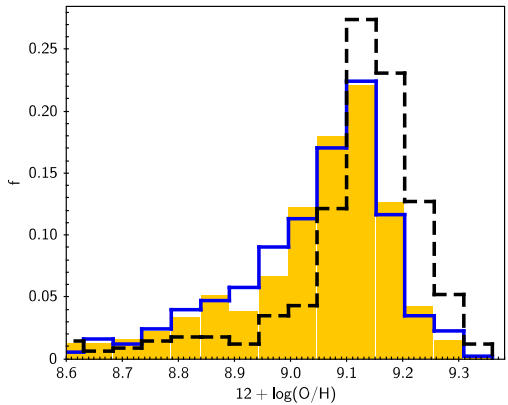}
\par\end{centering}

\caption{Metallicity distributions for unbarred (shaded histogram), weak-barred
(solid line) and strong-barred (dashed line) galaxies.}

\end{figure}

We also study the mass-metallicity relation (MZR; Lequeux et al.1979) of barred galaxies 
as a tool to study the effects of bars on the galaxy metallicity. In the local Universe, Tremonti et al. (2004) have confirmed the dependence of metallicity on stellar mass with high statistical signal.
Erb et al. (2006) has extended the study to high redshift finding a similar correlation, although
displaced to lower metallicity (Maiolino et al. 2007).
Fig. 12 shows the mass-metallicity relation for each galaxy type.
We also compare these with the results obtained of Tremonti et al. (2004), who studied mass-metallicity relation for a sample of 53000 star-forming galaxies from SDSS. 
It can be seen that our galaxy samples are more metallic, according to the results of Ellison et al. (2008, 2011). These authors found that mass-metallicity relation is modulated by the star formation rate. They suggest that the metal enhancement without an accompanying increase in the star formation activity may be due to a short lived phase of bar-triggered star formation in the past.
However, the most interesting point is that strong-barred objects show a relation which falls abruptly, with respect to ones to the other samples.
This tendency is more significant in low stellar mass galaxies. Nevertheless, for strong barred galaxies with $log(M_{*}/M_{\sun}) > 10.0$ there is not a clear fall in the metallicity, and also similar trends are observed for different samples.
 This fact could be indicating that prominent bars produce an accelerating effect on the gas processing and hence on the host galaxy evolution towards earlier morphological types.

 In addition, it can be seen that weak-barred and unbarred galaxies do not show significant differences in the metallicity. In a same direction, this behavior is reflected in the other galaxy properties: colors, star formation activity and stellar population. These findings could be indicating that weak bars do not produce noticeable changes in the galaxy properties and the effects on the physical characteristics begin to be felt when bar became prominent enough.

\begin{figure}

\begin{centering}
\includegraphics[scale=0.30]{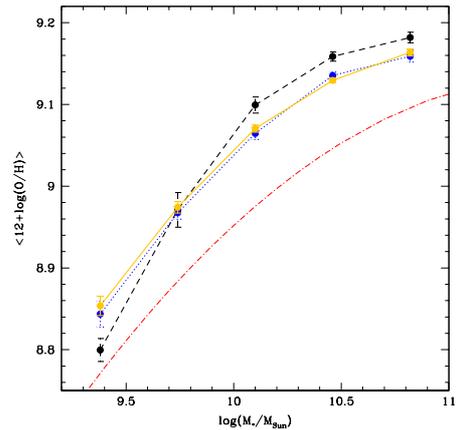}
\caption{Mass-metallicity relation for unbarred (solid line), weak-barred
(dotted line) and strong-barred (dashed line) galaxies.
The red dotted line represents the Tremonti et al. (2004) fit.}

\par\end{centering}

\end{figure}


\section{Summary and conclusions}

We have performed a statistical study of physical properties of 
barred galaxies in contrast with unbarred ones. Our analysis is
 based on a sample derived from SDSS release. 
We complemented these data with a naked-eye classification of 
a sample of face-on spiral galaxies brigther than $g = 16.5$ mag,  based 
on the presence of the bar, and taking into account the strength 
of the bar with respect to the structural properties 
of the host galaxies.  
With the purpose of providing an appropriate quantification
 of the effects of bars on host galaxies, we also constructed a suitable control sample of 
unbarred galaxies with the same redshift, $r-band$ magnitude,
 concentration index, bulge size parameter, and local environment 
distributions, following Alonso et al. (2013).

We can summarize the principal results of our analysis in the following conclusions:

\begin{enumerate}

\item We found 522 strong-barred, 770 weak-barred, and 3711 non-barred galaxies, 
which represents a bar fraction of 25.82$\%$, with respect to the full sample 
of spiral galaxies.
This fraction agrees with previous studies found by other authors by visual 
inspection of different galaxy samples from optical images (Nilson 1973, 
de Vaucouleurs et al. 1991, Marinova et al. 2009, Masters et al. 2010b, Alonso et al. 2013).

\item We observed that strong-barred galaxies show lower star formation activity and older stellar populations, with respect to weak-barred and unbarred disc objects.
We also found a significant fraction ($\approx20\%$) of strong-barred galaxies with older 
stellar population and low efficient star formation rate that have lenticular morphology (SB0 type). 
This result shows that, when S0 galaxies contain bar, it is usually a strong structure, 
in agreement with Aguerri et al. (2009).

\item We also studied the star formation activity and the age of stellar populations of galaxies as a function  of $log(M_{*})$ and concentration index, $C$, in barred galaxies with weak/strong bars, and in the control sample. 
We found that strong-barred galaxies show a systematically lesser efficient star formation 
activity and older stellar population for different stellar mass bins, and towards earlier 
morphology, with respect to the other samples of galaxies with weak and without bars.

\item We examined the color distributions of different samples studied in this work, and we found that there is a significant excess of strong barred host galaxies with red colors.
We also found that galaxies with strong bars are redder, for the whole concentration index range, with respect to their counterparts of weak-barred and unbarred disc objects.
In particular, for strong barred galaxies that belong to the minor peaks of the star formation and stellar population  distributions (see Fig. 3) these tendencies are clearly more significant, showing a high fraction of host galaxies with extremely red colors.
These findings suggest that bar perturbations have a considerable effect in modifying galaxy colors in the host galaxies, producing an acceleration of the gas processing, when bar became prominent enough.

\item The color-magnitude and color-color diagrams show that strong-barred galaxies are mostly grouped in the red region, while unbarred and weak-barred objects are more extended to the blue region. 
The positions in the color diagrams, could indicate the existence of an evolutive relation between the different considered galaxy type. In this scenario, an unbarred galaxy would begin to form a bar as a consequence of a gravitational disturbance in the disk. 
Then, matter would fall down to the center of the galaxy, making place to a weak bar which would become gradually more prominent while the inflow accumulates material in the center. 
At first, the weak bar would not be able to alter significantly the host characteristics, but then, when this structure is strong enough, it could affect significantly the galaxy properties.

\item We also explore the metallicity, which principally reflects the amount of gas reprocessed by the stars. 
It shows that galaxies with strong bars present an important excess towards high metallicity values, while unbarred and weak-barred disc objects have similar distributions.  
The mass-metallicity relation reflects that although unbarred and weak-barred galaxies are fitted by similar curves, strong-barred ones show a curve which falls abruptly.
It is more important in low stellar mass galaxies ($log(M_{*}/M_{\sun}) < 10.0$).  
This behaviour could be suggesting that prominent bars produce an accelerating effect on the gas processing, producing significant changes in the physical properties, also reflected in the evolutionary stages of the host galaxies.

\end{enumerate}

\begin{acknowledgements}
      This work was partially supported by the Consejo Nacional de Investigaciones
Cient\'{\i}ficas y T\'ecnicas and the Secretar\'{\i}a de Ciencia y T\'ecnica 
de la Universidad Nacional de San Juan.

Funding for the SDSS has been provided by the Alfred P. Sloan
Foundation, the Participating Institutions, the National Science Foundation,
the U.S. Department of Energy, the National Aeronautics and Space
Administration, the Japanese Monbukagakusho, the Max Planck Society, and the
Higher Education Funding Council for England. The SDSS Web Site is
http://www.sdss.org/.

The SDSS is managed by the Astrophysical Research Consortium for the
Participating Institutions. The participating institutions are the American
Museum of Natural History, Astrophysical Institute Potsdam, University of
Basel, University of Cambridge, Case Western Reserve University,
University of Chicago, Drexel University, Fermilab, the Institute for Advanced Study, the
Japan Participation Group, Johns Hopkins University, the Joint Institute for
Nuclear Astrophysics, the Kavli Institute for Particle Astrophysics and
Cosmology, the Korean Scientist Group, the Chinese Academy of Sciences
(LAMOST), Los Alamos National Laboratory, the Max-Planck-Institute for
Astronomy (MPIA), the Max-Planck-Institute for Astrophysics (MPA), New Mexico
State University, Ohio State University, University of Pittsburgh, University
of Portsmouth, Princeton University, the United States Naval Observatory, and
the University of Washington.
\end{acknowledgements}

\end{document}